# Control of silver-polymer aggregation mechanism by primary particle spatial correlations in dynamic fractal-like geometry

Gaetano Campi,[1] Alessandra Mari,[2] Augusto Pifferi,[1] Heinz Amenitsch[3], Michela Fratini[4] and Lorenza Suber[2]

[1]CNR- Istituto di Cristallografia, Via Salaria, Km 29.300, Monterotondo Stazione, (RM), I-00015, Italy
[2]CNR- Istituto di Struttura della Materia, Via Salaria, Km 29.300, Monterotondo Stazione, (RM), I-00015, Italy
[3]Institut of Biophysics and Nanosystems Research, Austrian Academy of Sciences, Schmiedlstrasse 6, 8042 Graz, Austria
[4]CNR- Istituto di Fotonica e Nanotecnologie, Via Cineto Romano 42, 00156 Roma, Italy.

**ABSTRACT:** Silver nanocrystals have been prepared by reacting silver nitrate with ascorbic acid in aqueous solution containing a low concentration of a commercial polynaphtalene sulphonate polymer (Daxad 19). Various crystalline morphologies have been obtained simply by tuning the reaction temperature. We have investigated the nanoparticle formation mechanism at three different temperatures by in situ and time resolved Small Angle X ray Scattering measurements. By modeling the scattering intensity with interacting spherical particles in a fractal-like polymer-Ag matrix, we found signatures of nucleation, growth and assembly of primary particles of about 15-20 nm. We observed how the time evolution of both spatial correlations between primary particles and the dynamic fractal geometry of the polymer-Ag matrix could influence and determine both the aggregation mechanism and the morphology of forming nanostructures in solution.

## 1. Introduction

Growing and renewed interest in nanoparticle research field has been recently placed on metal nanoparticles due to their unique physicochemical properties and several potential applications e.g. in selective catalysis [Lewis, 1993], photonics [Maier, 2001], optoelectronics [Kamat, 2002] and biomedicine [Nicewarner, 2001]. Toward controlled nanoparticle synthesis, a widely used method consists of using capping molecular agents such as organic surfactants and polymers in wet chemical reactions to (template) tailor the size and shape of forming nano-composited architectures [Fendler, 1996]. Since the intrinsic properties of nanostructures are highly influenced by their shape, size and structure [Bonnemann, 2001], several efforts have been made for investigating the interplay of their formation mechanism and the causes at the base of the different morphologies. In this context, understanding clusters nucleation and growth as well as their assembly, appears to be essential for both a practical design strategy in the synthesis of nanostructured systems [Brechignac, 1998; Kaiser, 2004] and for tuning their properties.

In an aqueous chemical synthesis approach, size, shape and structure of nanoparticles can be determined by tuning various experimental parameters such as, to name a few, reaction temperature and time, pH, reagent concentrations and addition rate. Thus, the study of how the final morphology depends on the above-mentioned parameters is important to the understanding of the fundamental growth mechanisms involved in pattern formation. Despite huge progress in the practical defining of the synthesis routes, the mechanism of the nanoparticle formation as well as their shape control is not well understood, in part due to the fact that the interplay of the above mentioned experimental parameters, that is the interactions taking place during the particle formation in the solution, are quite complex.

We recently exploited a silver particle synthesis utilizing silver nitrate as salt, nitric acid to set acidic condition, ascorbic acid as reducing agent and polynaphtalene sulphonate polymer, termed Daxad 19, as particle stabilizer in aqueous phase [Suber, 2005, 2009, 2010]. In this system, we have found that different silver-polymer morphologies can be selectively grown, through the control of only one parameter, the reaction temperature, while all other experimental parameters such as pH, concentrations of all species in solution and the addition rate of reducing ascorbic acid, are fixed. This emerged simplicity led us to investigate their formation mechanism simply by tuning the temperature in the chemical reaction; in particular, in this paper, we report the syntheses carried out at 20, 40 and 60 °C at which rods-like, hexagonal-like and chain-like polymer-Ag composites are obtained respectively.

In order to understand which parameters are in turn switched on by thermal energy for controlling size and shapes of the final morphologies, we investigated the kinetic of $Ag^+$ reduction by using time resolved Small Angle X-ray Scattering at the selected temperatures. We used an experimental setup that allowed us to control the $Ag^+$ to $Ag^0$ reduction time obtaining characteristic times of the order of about 300 seconds, well suited to be studied by synchrotron in situ SAXS technique. At the end of the reactions at the different temperatures, we observed the formation of crystalline silver colloids; subsequently, the SAXS data analysis suggested us that these colloids form through nucleation, growth and aggregation of fractal silver-polymer primary particles assisted by long range spatial correlations, *LRC*. Since these *LRC* correlations control the magnitude as well as the range of the interaction between the colloidal particles, they are believed to play a relevant role in the formation of the final morphologies as well as in tuning their properties [Murray et al. 2000].

In nanoparticles wet chemical synthesis, polymers are usually used to hinder particle-particle aggregation and to control interparticle distances; for example, Rotello [Frankamp, 2002] and coworkers used poly(amido-amine) PAMAM dendrimers of different generation in order to control the gold interparticle distance; at the same time the gold nanoparticles were functionalized with carboxylic acid terminal groups, in order to provide the driving force for assembly. In our case, Daxad, with its sulphonic groups [Suber, 2005] plays a double role since first it coordinates $Ag^+$ ions and then, as soon as the Ag° primary particles are formed, correlates them designing fractal geometries.

A wide variety of physical systems present fractal geometry [Pietronero, 1986] with spatial correlations [Stanley, 1993]; the understanding and characterization of these correlations represent a challenge in theoretical physics in last decades and although self-organizing criticality [Aharony, 1989] has been the preferred description for these phenomena, new, more complex and unexpected insights enrich this field of research [Amaral, 1998, Fratini, 2010; Cavagna, 2010]. In this view, our experimental data describe the interplay of spatial correlations between primary particles and the isothermal dynamic arrangement of the fractal-like polymer-Ag matrix in the early stage of particle formation. We hope this work could represent an experimental contribution to the topical double handed issue pertaining both to technology, relative to controlled fabrication of nanoparticles, and to basic chemical physics, dealing with the formation and evolution of order in aggregation processes.

## 2. Experimental

**2.1** Materials. Silver nitrate and ascorbic acid, purchased from Aldrich, were of the highest purity grade. $H_2O$ Plus for HPLC and $HNO_3$ 69.5 wt% were purchased from Carlo Erba. Daxad 19 (the sodium salt of a polynaphthalene sulfonate-formaldehyde condensate, having 80 kD av mol wt), indicated in the following as Daxad, was obtained from Hampshire Chemical Company. Elemental analysis (wt %) of the product gave C: 42.42, H: 3.44, S: 6.20.

**2.2** Sample preparation and characterization. AgNO3 (1.02 g, 6.0 mmol) was dissolved in 30 mL of water and the solution was thermostated at 40 ± 2 °C. A 0.178 g of Daxad and 3.1 mL of HNO3 were added under mechanical stirring. The procedure was completed in two minutes. Then ascorbic acid (1.2 g, 6.8 mmol), dissolved in 4.0 mL of water, was added at a 0.5 mL/min rate. After 1 hour stirring at the above temperature, the flask was cooled to 20 °C. After settling off the light grey precipitate, the mother liquors were siphoned, the precipitate washed three times with deionised water and dried under vacuum. The same procedure, at the different temperatures, was used for the samples prepared at 20 and 60°C.

**2.3 Data Analyses and Instruments**

SEM measurements were performed at 20 kV with a SEM JEOL 6400. SAXS measurements were performed on the Austrian SAXS beamline at ELETTRA, Trieste [Amenitsch et al., 1997]; we set the camera to a sample detector distance of 1.5 m and operated at photon energy of 8 KeV covering the *q* range between 0.07 and 1.7 $nm^{-1}$. We used a thermostated batch reactor apparatus consisting of a glass flask in which the reaction takes place, a remote controlled syringe that allowed us to add 4 mL of ascorbic acid solution at the fixed rate of 0.5 mL/min into the reaction solution and a peristaltic pump that continuously flows the solution mixture in a 1.5 quartz capillary through a closed circuit; the pumping rate was set to 20mL/min in order to change all the tubing (1m x 2mm) volume in less than 10 seconds, avoiding particle deposition on the walls. Two dimensional SAXS patterns were thus regularly recorded on a CCD camera with a time resolution of 0.9 sec. Data have been treated involving detector corrections for flat field response, spatial distortion, dark current of the CCD and normalization by the incident flux; water flowing was also measured in order to asses and subtract the background from the data. The resulting two-dimensional images were integrated to obtain 1d pattern of normalized intensity versus scattering vector *q*.

**3. Results and discussion**

Ag colloidal particles are formed by reduction of $Ag^+$ with ascorbic acid in an acidic aqueous solution of a 0.5 % w/w commercial polynaphtalene sulphonate polymer termed Daxad. SEM images of particles synthesized at T=20 °C, T=40 °C and T=60 °C are shown in Fig. 1. Irregular rods-like structures are produced at T=20 °C, stable single particles at 40°C and from the reaction performed at T=60 °C chain-like particle structures are formed. XRD powder measurements (see the insets in SEM images of Fig. 1) show that the rods, particles and particle chains are made of silver crystalline subunits of about 20 nm, as obtained from Scherrer analysis of diffraction peaks. Electrophoretic measurements indicate that the platelets, particles and particle chain structures are covered, as expected, by the polymer [Suber, 2005]. In order to understand how these crystalline subunits form and aggregate at the different temperatures, we performed in situ and time resolved synchrotron SAXS measurements.

Aqueous solution of $AgNO_3$ and Daxad was prepared before the addition of ascorbic acid at $t=t_o=10s$. Time evolution of SAXS intensity at different time intervals is shown in Fig. 2 for the three investigated temperatures. For each temperature, the SAXS profiles present the following main features: *i)* the scattered intensity gradually increases in the low *q* region indicating the nucleation and growth, *NG*, of Ag-polymer particles; *ii)* a change in the profile slopes, in the high *q* region; *iii)* the development of interference oscillations particularly evident at T=40 °C.

In order to describe and quantify these peculiar aspects of our experimental data we assume a theoretical model consisting of two components. The first component, $I_B(q)$, takes into account the power law behaviour with exponent $P_E$ predominating at high *q* and is ascribed to like-fractal clusters in the solution; the second one, $I_P(q)$, describes highly dispersed particles interacting via hard-sphere potential. Thus the model *I(q)* can be written as

$$I(q) = I_B(q) + I_P(q) \tag{1}$$

where

$$I_B(q) = P_C q^{-P_E} \tag{2}$$

with $P_C$ is a constant and $P_E$ Porod exponent, and

$$I_P(q) = CnS(q, R_{HS}, \eta) \int_0^\infty P(R)[V(R)\Phi(R)R]^2 \, dR \tag{3}$$

takes into account the formation of *n* spherical primary particles with radius *R*; *C* is a constant independent of *q* and *R*; *V(R)* and *Φ(q,R)* are the volume and the form factor of the single spherical particle, respectively. In order to account for the polydispersivity of the particles, the intensity has been integrated over a log-normal distribution of *R*, *P(R)* given by

$$P(R) = \frac{1}{R\sigma\sqrt{2\pi}} \exp - \frac{(\ln R - \rho)^2}{2\sigma^2}, \text{ with } \int_0^\infty P(R)dR = 1 \tag{4}$$

Finally, the scattering oscillations have been modelled in the monodisperse approximation of hard spheres with radius $R_{HS}$ and volume fraction $\eta$ calculated with the Percus-Yevick equation [Percus, 1958]. We used the trust-region-reflective algorithm [Coleman, 1994], in order to fit the data and estimate the time dependence of the $P_C$ constant and the power exponent $P_E$ of the background term in $I_B(q)$, the number density *N(=Cn)*, the mean radius *R* and the dispersivity $\sigma$ of the *log-normal* primary particles distribution size *P(R)*, the hard sphere radius $R_{HS}$ and volume fraction $\eta$ in the Percus-Yevick approximation. We have verified the reliability of refined parameters, by inspection of their confidence intervals [Coleman, 1994] from which we obtained the reduced standard deviation, *rds*, of all parameters. The time evolution of all *rds*, shown in Fig. 1a, 1b and 1c of SI, specify the range of validity and the limitations of our model. In particular, we find that the *rds* on the refined parameters R, σ, N, (describing the $I_p(q)$ term) exceeds 20% before the instants $t^*_{20}$=50s, $t^*_{40}$=20s and $t^*_{60}$=15s at T= 20 °C, 40 °C and 60 °C respectively; at the same time, the *rds* $\Delta P_E/P_E$ of the $P_E$ exponent (describing the $I_B(q)$ term) is found to be generally quite small and acceptable. This indicates at each temperature, before the t* instants, the scattered intensity is largely dominated by the background, $I_B(q)$, preventing us from estimating the particle size distribution. Thus, before $t_0$, the SAXS intensity can be related to a fractal system made of large clusters with dimension $P_E$ and size $\xi$ with $q\xi \gg 1$ in the measured $q$ range [Freltoft, 1986]. As soon as the reduction process by ascorbic acid takes place ($t=t_0$), $Ag^0$ clusters nucleate and grow, as described by the time evolution of their total volume, $V_p$, given by

$$V_p = N \int_0^\infty P(R)V(R)dR \qquad (5)$$

Fig. 3 shows the plot of $V_p/V_{final}$ as a function of the time, where $V_{final}$ represents the volume values of primary particles at the end of the reduction process. Considering the time when $V_p$ reaches the half of the final value $V_{final}$, namely $t_{1/2}$, as the characteristic time of the *NG* process, we find that nucleation and growth rate increases with the temperature, as expected. Furthermore, larger particles are obtained for higher *NG* rates, as can be deduced by Tab.1. Surprisingly, this appears to be in contrast with the results of Abecassis and co-workers [Abecassis, 2007]; they synthesize gold nanoparticles controlling the *NG* rate using a different reducing agent, finding smaller final particle size at higher *NG* rate. This contrast could be ascribed to the different reaction times: while their reduction time is of the order of the seconds, in our case the *NG* period is much longer, particularly at lower temperatures. Thus, scale-time of the reaction seems a further relevant parameter to be considered in kinetics studies of nanoparticle formation.

After the nucleation and growth takes place, diffraction peaks arise, indicating spatial correlations on different volume fractions and distances as a function of both the temperature and the time (see Fig. 2 and Fig. 4). Numerous experimental small-angle neutron or X-ray scattering studies in aqueous solutions have identified peaks in the distribution of the scattered intensities. We cite here relevant works in which low-angle diffraction peaks growing and moving towards lower $q$ values have been associated to the particles growth [Langmayr, 1992], cluster-cluster aggregation [Sciortino, 1995], spinodal decomposition [Carpineti, 1992], gelation of emulsions [Bibette, 1992]. Furthermore, the appearance of peaks in the structure factor at high wave number, $q$, has been reported in several recent works dealing with aggregation of compact clusters [Puertas, 2003, Cates, 2004; Wilking, 2006].

We find dynamic interference oscillations (Fig. 4) indicating that particles, on average, are correlated on distances defined by $2\pi/q_m = R_{HS}$ where $q_m$ is referred to the main peak position in the pattern and $R_{HS}$ is the hard sphere radius in the Percus-Yevick approximation. These correlations are stronger at 40 °C (Fig. 2, Fig.4) where the ordering turns on after an incubation time of about

10s and involves all the available volume fraction of 0.3 on nearly constant distances (≈24nm) when t=$t_{\eta max}$=43s (see Fig. 4); this behavior is consistent with a nucleation regime (I) typical of cooperative phenomena. Subsequently, for $t_{\eta max}$<t<$t_{co}$, the time evolution of spatial correlations (see Fig. 2a) shows an ordering regime (II) characterized by a nearly linear shifting, $\alpha_{II}(t-t_{\eta max})$, of the correlation length; finally at t>$t_{co}$ the system enters in a regime (III) where the increase of the spatial correlations does not saturate, increasing with a power law growth, $\alpha_{III}(t-t_{co})^\gamma$, with exponent $\gamma$ = 0.026 ± 0.004 (see inset in Fig. 4a). Temperature changes leads to weaker correlations, as shown in Fig. 4. More specifically, at T=20 °C we can appreciate weak correlations developing on increasing distances from 24 up to 70 nm and on volume fractions that become negligible at about t=100s; even weaker correlations are found at T=60 °C, with nearly constant distances of 70 nm and volume fractions going to negligible values for t=100s.

The arising and development of these correlations during the reduction process, come out from the polymer matrix arrangement; this latter, thus, can be seen as playing the double role of ordering and protecting agent, and could drive the ordered aggregation of primary particles. In order to characterize and get deeper insight on the polymer role in the aggregation process, we consider the time evolution of the Ag-polymer background matrix in the solution, described by the power law background component $I_B(q)$ in Eq. 1. At the beginning of the $Ag^+$ reduction (t>$t_0$) $P_E$ assume low values, less than 2, typical of diffusion limited regimes. As the reaction goes on and the colloid forms, $P_E$ increases and reaches its steady value of about 2.1 at T=20 °C and 2.5 at T=60 °C, typical of reaction limited aggregation regimes (Fig. 5) [Jullien, 1987]. On the other hand, at T=40 °C $P_E$ reaches values around 3.2; since $P_E$ values from 1 to 3 indicate *mass fractal* dimension, at T=20 °C and 60 °C we have mass fractal $Ag^0$-polymer clusters in the colloid; as $P_E$ exceeds 3 the system undergoes a mass-surface fractal transition. Thus, after $t_{co}$, we find more compact objects only in the colloid formed at T=40 °C where a mass-surface fractal transition occurs [Campi, 2010].

The long ranged spatial correlations persisting at this temperature appear not saturating after the *mass-surface* fractal transition at $t_{co}$, and increase following a weak power law (inset in Fig. 4a); this behavior can be ascribed to a *like-coarsening* regime assisting aggregation mechanism at 40 °C.

Both the power law of spatial correlations and their surprising long length in comparison with the particle size, found at T=40 °C, could lead to hypothesize the occurrence of a critical point in the system at this temperature. Indeed, it has been found that LRC develop at certain critical points in many systems in non-equilibrium state, playing an important role in the formation of self-organized structures [Giglio, 1981; Chatè, 1995]. Nevertheless, the arising and development of long range correlations in complex systems still constitute an unsolved theme; for examples, power law scaling occurring also in the absence of critical dynamics has been proposed and investigated in biological or social systems composed by units with complex evolving structure [Amaral, 1998]; at the same time, possible mechanism explaining the influence of correlations on the morphology of evolving fractal structures could be the correlated clustering in networks formation and evolution [Maske, 1998].

Beyond this complexity, in this work we describe a scenario where primary nanoparticles form and aggregate through fractal polymer building units that reorganize themselves during the reduction process. Thus, these primary particles are capped in a matrix provided by the polymer; as they nucleate and grow, the matrix is able to connect and correlate them on temperature dependent spatial distances, as probed by the time evolution of the structure factor $S(q,R_{HS},\eta)$. More specifically, $Ag^+$ ions, able to coordinate to the sulfonic groups of the polymer, act as connecting nodes that link the sulfonic groups of different polymer building units giving rise to a network that evolves from *mass fractal* type to *surface fractal* type at T=40 °C, while remains mass fractal at 60 °C and 20 °C. In other words, the complex polymer-silver interactions developing during the reaction seem to provide a kind of "*dynamical template*" changing with the reaction temperature and leading to different fractal aggregation types and different morphologies.

## 4. Summary

We have used time resolved Small-Angle X ray Scattering at three different temperatures for investigating the mechanism behind the formation and aggregation of crystalline colloidal $Ag^0$ particles by $Ag^+$ reduction with ascorbic acid in an aqueous acidic solution of a polynaphtalene sulphonate polymer (Daxad). Data analysis suggest that the nucleation, growth and aggregation of primary particles of 15-20 nm in diameter is assisted by the development of long range liquid-like spatial correlations; these correlations are driven by Ag-polymer interactions that depend on the temperature and provide a like-fractal "*dynamic template*" in the solution. Thus, the temperature can be seen as the early cause that rules the assembly of primary particles: stable hard silver nanoparticles of 120 nm are obtained at T=40 °C while friable larger structures and like-chain structures are formed at lower (20 °C) and higher temperature (60 °C), respectively. In conclusion, our data show as the extent of spatial correlations in variable fractal geometries, during nanoparticle formation, constitutes a key point for understanding the different morphologies of the emerging patterns.

**Acknowledgements**

We thank Dr Patrizia Imperatori for XRD analysis and Antonello Ranieri and Andrea Notargiacomo for their technical support. Alessandra Mari thanks the CNR Ricerca Spontanea a Tema Libero project (RSTL 087.008) for financial support.

# References

L. N. Lewis, **Chemical catalysis by colloids and clusters**, *Chem. Rev.* **93**, 2693 (1993).

S. A. Maier, M. L. Brongersma, P. G. Kik, S. Meltzer, A. A. G. Requicha, H. A. Atwater, **Plasmonics—A Route to Nanoscale Optical Devices,** *Adv. Mater.* **13**, 1501 (2001).

P. V. Kamat, **Photophysical, Photochemical and Photocatalytic Aspects of Metal Nanoparticles**, *J. Phys. Chem. B* **106**, 7729 (2002).

S. R. Nicewarner-Peña, R. G. Freeman, B. D. Reiss, L. He, D. J. Peña, I. D. Walton, R. Cromer, C. D. Keating, M. J. Natan, **Submicrometer Metallic Barcodes**, *Science* **294**, 137 (2001)

J.H. Fendler, **Self-Assembled Nanostructured Materials**, *Chem.Mater*. 8, 1616-1624 (1996)

Bonnemann H, Richards RM: **Nanoscopic metal particles- Synthetic methods and potential applications.** *Eur J Inorg Chem* 2001, **10:**2455-2480.

C. Brechignac, P. Cahuzac, F. Carlier, M. Defrutos, A. Masson, C. Mory, C. Colliex, B. Yoon, **Size effects in nucleation and growth processes from preformed soft-landed clusters,** *Phys. Rev. B*, **57**, R2084-R2087 (1998)

B. Kaiser, B. Stegemann, **Cluster Assembled Nanostructures: Designing Interface Properties**, *ChemPhysChem*, **5**, 37-42 (2004)

L. Suber, I. Sondi, E. Matijević and Dan V. Goia, **Preparation and the mechanism of formation of silver particles of different morphologies in homogeneous solutions**, *J. Colloid Interface Sci.*, **288**,489-495 (2005)

L. Suber, G. Campi, A. Pifferi, P. Andreozzi, C. La Mesa, H. Amenitsch,| R. Cocco, and W. R. Plunkett: **Polymer-Assisted Synthesis of Two-Dimensional Silver Meso-Structures**, *J. Phys. Chem. C*, **113:** 11198–11203 (2009)

L. Suber, W. R. Plunkett, **Formation mechanism of silver nanoparticle 1D microstructures and their hierarchical assembly into 3D superstructures,** *Nanoscale*, 2(1), 128-33 (2010)

C. B. Murray and C. R. Kagan, and M. G. Bawendi, **Synthesis and characterization of monodisperse nanocrystals and close-packed nanocrystal assemblies**, *Annual Review of Materials Science* **30**, 545-610 (2000)

B. L. Frankamp, A. K. Boal, and V. M. Rotello, **Controlled Interparticle Spacing through Self-Assembly of Au Nanoparticles and Poly(amidoamine) Dendrimers**, *J. Am. Chem. Soc.*, **124**, 15146-15147 (2002)

L. Pietronero and E. Toscatti, **Fractals in Physics**, edited by (North-Holland, Amsterdam, 1986)

H.E. Stanley, S.V. Buldyrev, A.L. Goldberger, S. Halvin, C.-K. Peng and M. Simons, **Long-range power-law correlations in condensed matter physics and biophysics,** *Physica A*, **200**, 4-24 (1993)

A. Aharony and J. Feder, **Fractals in Physics***,* Proceedings of a Conference, Vence, France (North Holland, Amsterdam, 1989)

L. A. N. Amaral, S. V. Buldyrev, S. Havlin, M. A. Salinger, and H. E. Stanley, **Power Law Scaling for a System of Interacting Units with Complex Internal Structure**, *Phys. Rev. Lett.*, 80, 1385-1388 (1998)

M. Fratini, N. Poccia, A. Ricci, G. Campi, M. Burghammer, G. Aeppli & A. Bianconi, **Scale-free structural organization favoring high temperature superconductivity**, *Nature*, **466**, 841-844 (2010)

A. Cavagna, A. Cimarelli, I. Giardina, G. Parisi, R. Santagati, F. Stefanini and M. Viale, **Scale-free correlations in starling flocks,** *Proc Natl Acad Sci USA*, **107**, 11865–11870 (2010)

H. Amenitsch, S. Bernstorff, M. Kriechbaum, D. Lombardo, H. Mio, M. Rappolt and P. Laggner, **Performance and first results of the ELETTRA high-flux beamline for small-angle X-ray scattering**, *Journal of Applied Crystallography,* **30,** 872-876 (1997)

Coleman, T.F. and Y. Li, "**On the Convergence of Reflective Newton Methods for Large-Scale Nonlinear Minimization Subject to Bounds**," *Mathematical Programming*, Vol. 67, Number 2, pp. 189-224, 1994.

J.K. Percus, G.J. Yevick, **Analysis of Classical Statistical Mechanics by Means of Collective Coordinates**, *J. Phys. Rev.* **110**,1-13 (1958)

G. Campi, A. Mari, H. Amenitsch, A. Pifferi, C. Cannas and L. Suber, **Monitoring early stages of silver particle formation in a polymer solution by in situ and time resolved Small Angle X ray Scattering**, *Nanoscale*, **2**, 2447-2455 (2010)B.

T. Freltoft, J. K. Kjems, S. K. Sinha, **Power law correlations and finite size effects in silica particles aggregates studied small-angle neutron scattering**, *Phys. Rev. B*, 33, 269-275 (1986)

Abécassis, F. Testard, O. Spalla, P. Barboux, **Probing in situ the Nucleation and Growth of Gold Nanoparticles by Small-Angle X-ray Scattering**, *Nano Lett.*, **7**, 1723 (2007)

F. Langmayr, P. Fratzl and G. Vogl, **Volume fraction dependence of the structure function in Al-Ag**, *Acta metal. Mater.* **40**, 3381 (1992)

F. Sciortino, A. Belloni, and Piero Tartaglia, **Irreversible diffusion-limited cluster aggregation: The behavior of scattered intensity**, *Phys. Rev. E*, **52**, 4068 (1995)

M. Carpineti and M. Giglio, **Spinodal Type Dynamics in Fractal Aggregation of Colloidal Clusters**, *Phys. Rev. Lett.*, **68**, 3327 (1992)

J. Bibette, T. G. Mason, Hu Gang, and D. A. Weitz, **Kinetically Induced Ordering in Gelation of Emulsions**, *Phys. Rev. Lett.*, **69**, 981 (1992)

A. M. Puertas, M. Fuchs, and M. E. Cates, **Simulation study of nonergodicity transitions: Gelation in colloidal systems with short-range attractions**, *Phys. Rev. E* **67**, 031406 (2003).

M. E. Cates, M. Fuchs, K. Kroy, W. C. K. Poon and A. M. Puertas*,* **Theory and simulation of gelation, arrest and yielding in attracting colloids**, *J. Phys. Condens. Matter* **16**, S4861 (2004).

J. N. Wilking, S. M. Graves, C. B. Chang, K. Meleson, M.Y. Lin, and T. G. Mason, **Dense Cluster Formation during Aggregation and Gelation of Attractive Slippery Nanoemulsion Droplets**, *Phys. Rev. Lett.* **96,** 015501 (2006)

R. Jullien, R. Botet, R. Jullian, **Aggregation and Fractal Aggregates,** World Scientific Publishing Co Pte Ltd, ISBN: 9789971502485 (1987)

M. Giglio, S. Musazzi and U. Perini, **Transition to Chaotic Behavior via a Reproducible Sequence of Period-Doubling Bifurcations**, *Phys. Rev. Lett.*, **47**, 243-246 (1981)

H. Chatè, G. Grinstein and L. H. Tang, **Long-Range Correlations in Systems with Coherent (Quasi)periodic Oscillations**, *Phys. Rev. Lett.*, **74**, 912-915 (1995)

H. A. Maske, J. S. Andrade, M. Batty, S. Havlin, H. E. Stanley, **Modeling Urban Growth Patterns with Correlated Percolation**, *Phys. Rev.E*, **58**, 7054-7062 (1998)

**Tables:**

|  | R (nm) | σ | $t_{1/2}$ (s) |
|---|---|---|---|
| T = 20 °C | 6.7 ± 1.1 | 0.63 ± 0.09 | 196 |
| T = 40 °C | 9.4 ± 0.7 | 0.51 ± 0.05 | 110 |
| T = 60 °C | 9.7 ± 1.0 | 0.47 ± 0.05 | 30 |

**Tab.1** Size R (radius), dispersivity σ, and NG rate $t_{1/2}$, at the different temperatures. The R and s values are the steady values of radius and dispersivity reached at the end of the reduction process, while their errors correspond to their confidence interval in the fitting procedure.

**Figures:**

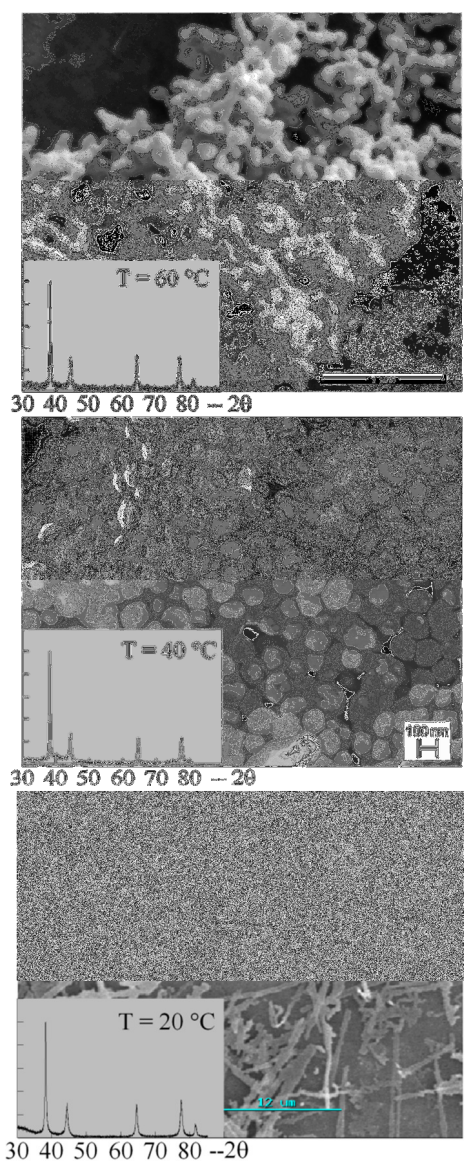

**Figure 1:** SEM micrograph and XRD pattern of final silver particles obtained at T=20, 40 and 60 °C. The size bars correspond to 1000 nm, 100nm and 12 µm at 60, 40 and 20 °C respectively .

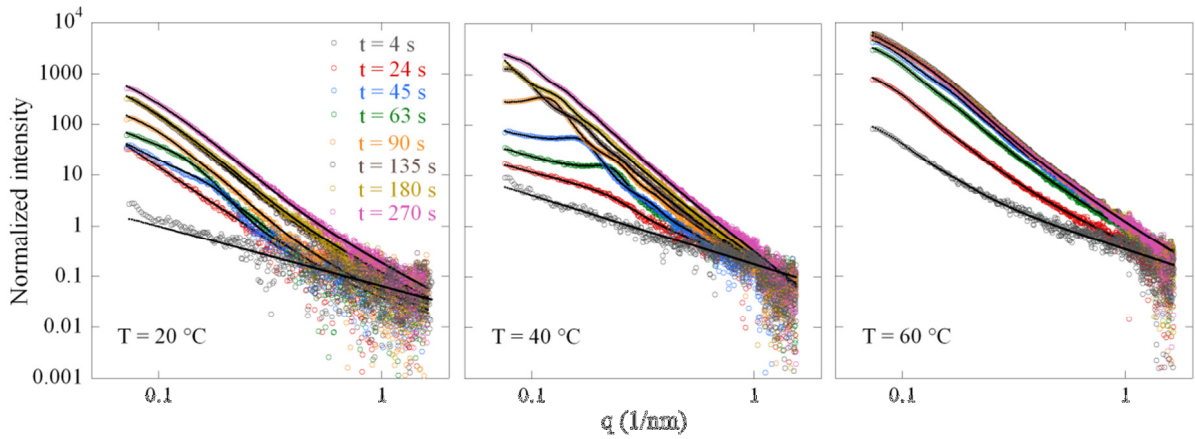

**Figure 2:** SAXS normalized profiles (open circles) collected at the time intervals indicated, at T=20 °C and T=40 °C and T=60 °C; solid lines show the best-fitted curves calculated by Eq. 1.

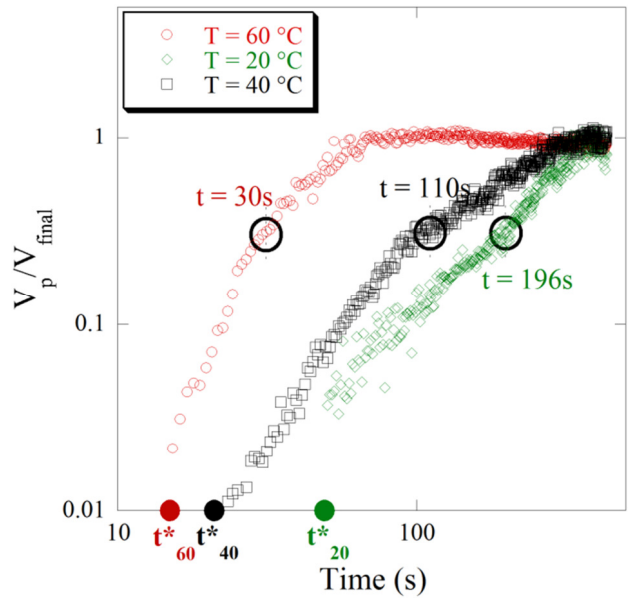

**Figure 3:** Representation of the *NG* process by time evolution of $V_p/V_{final}$ at the three investigated temperatures; $V_p$ is the total volume of particles while $V_{final}$ is the steady total volume of particles reached during the reduction process. The $t_{1/2}$ instants (see text) are indicated by the open circles, while the full circles on the time axis represent the t* instants (see text).

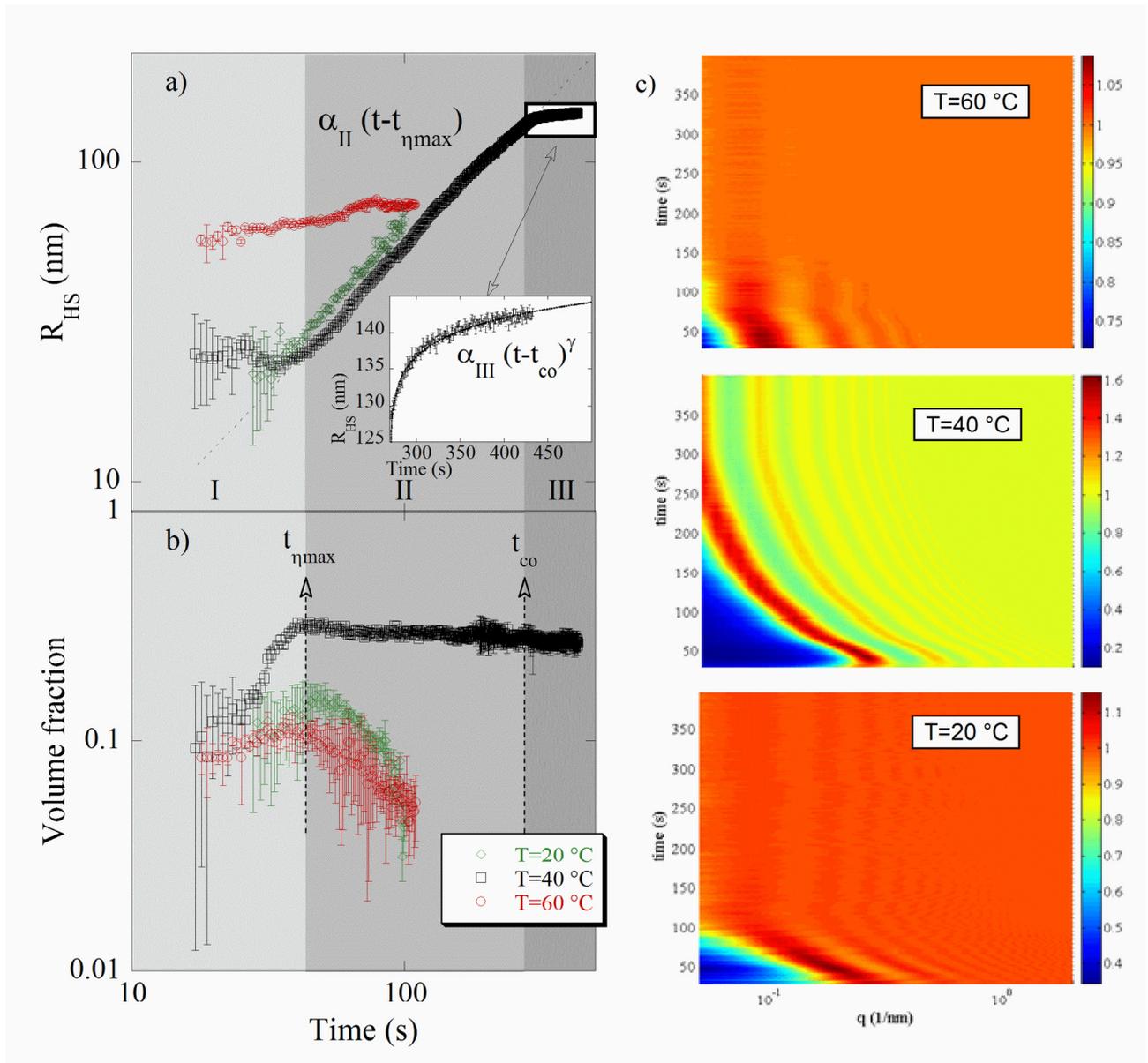

**Figure 4:** Time evolution of (a) $R_{HS}$ and (b) volume fraction $\eta$. Correlations can be appreciated at about 10 seconds after the starting of introduction of ascorbic acid at T=40 °C and T=60 °C, while they start after further 10 seconds at lower temperatures (T=20 °C). At T=40°C the spatial length correlation $R_{HS}$ reaches about 140 nm; at the other two temperatures its maximum value is found to be about 70 nm. At $t_{\eta max}$=43s, (dashed line in the bottom panel) the correlated volume fraction reaches its maximum value; subsequently, it remains nearly constant at T=40 °C while it decreases, assuming negligible values for t approaching 100s at T=20 °C and T=60 °C. In the inset in a) is highlited the power law behaviour of correlations occurring at $t_{co}$ (40 °C). The Percus-Yevick structure factors at the three temperatures are shown on the left as coloured maps in *time-q* plane.

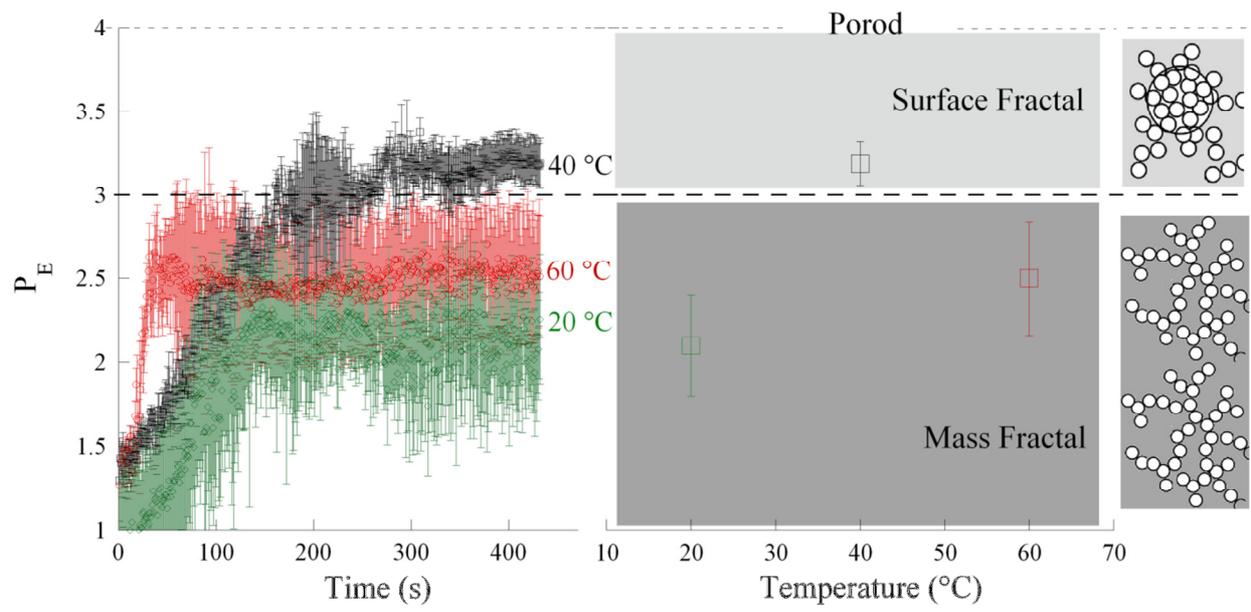

**Figure 5:** (Left panel) Fractal background dimension given by the exponent $P_E$ in the power law of Eq. 2., at the three different temperatures; confidence intervals from optimization procedure are indicated with error bars. (Right panel) Final steady values reached by $P_E$ at the different temperatures; the darker area ($1<P_E<3$) indicates mass fractal regime, while the area corresponding to the $3<P_E<4$ range denotes the surface fractal regime.

# Supplementary information

We have verified the reliability of refined parameters by calculating their confidence intervals. The reduced standard deviations (*rsd*) $\Delta R/R$, $\Delta\sigma/\sigma$, $\Delta N/N$, of the refined parameters $R$, $\sigma$ and $N$ are plotted as a function of the time in Fig. 1a, 1b, 1c, respectively. For $t<t_0$, $\Delta N/N$, $\Delta R/R$, $\Delta\sigma/\sigma$, are quite large; as the reaction takes place, the *rsd* on the refined parameters $R$, $\sigma$, $N$, (describing the $I_P(q)$ term) exceeds 20% before the instants $t^*_{20}=50s$, $t^*_{40}=20s$ and $t^*_{60}=15s$ at 20 °C, 40 °C and 60 °C respectively. In this time ranges, $\Delta P_E/P_E$ generally assume small reasonable values (Fig. 2), although we note large $\Delta P_E/P_E$ values at the beginning of the reaction at T=20 °C due to noisy scattering in the high $q$ region. These observations indicate that the contribution $I_p(q)$ to the model of eq.1 is almost negligible before the starting of the reactions; plus, considering *rsd* < 0.2, as a reasonable threshold for the reliability of refined parameters, the model of Fig.1 starts to work properly at the t* instants.

The model of eq. 1 has been derived considering the interference effects due to monodisperse hard spheres; in Fig. 3 we can check the reliability of this approximation. We observe that the standard deviation of both $R_{HS}$ and $\eta$ assume sufficiently small values as the reaction takes place and goes on at T=40 °C and T=20 °C, while large fluctuations of $\eta$ take place at 60 °C also at t>50s.

**Figures**

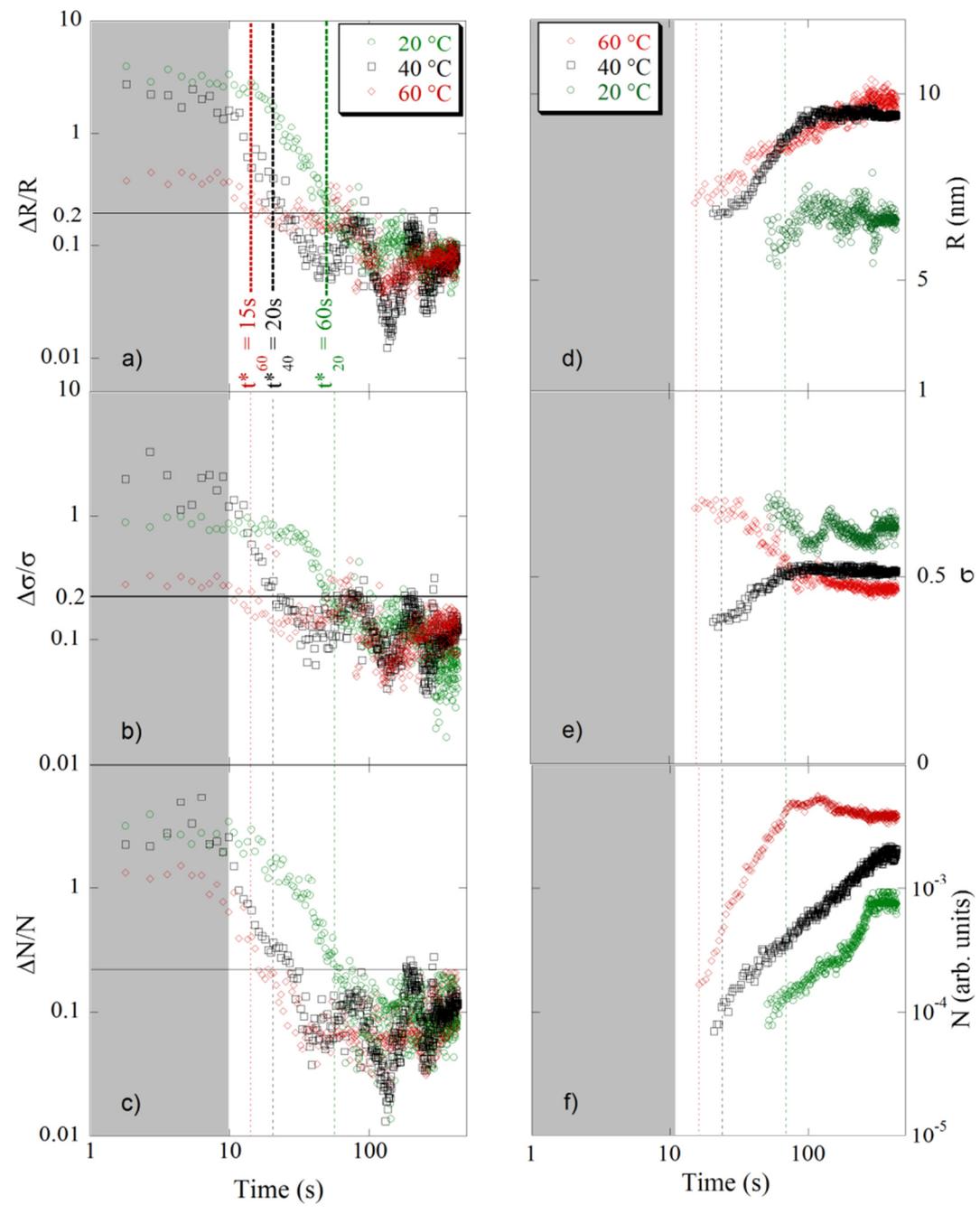

**Figure 1**: a) *ΔR/R*, b) *Δσ/σ*, c) *ΔN/N* d) *R*, e) *σ* and f) *N* plotted as a function of the time.

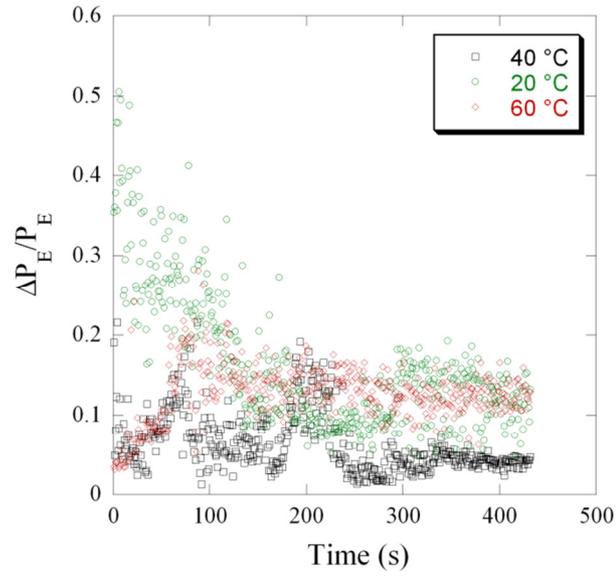

**Figure 2**: $\Delta P_E/P_E$ as a function of the time.

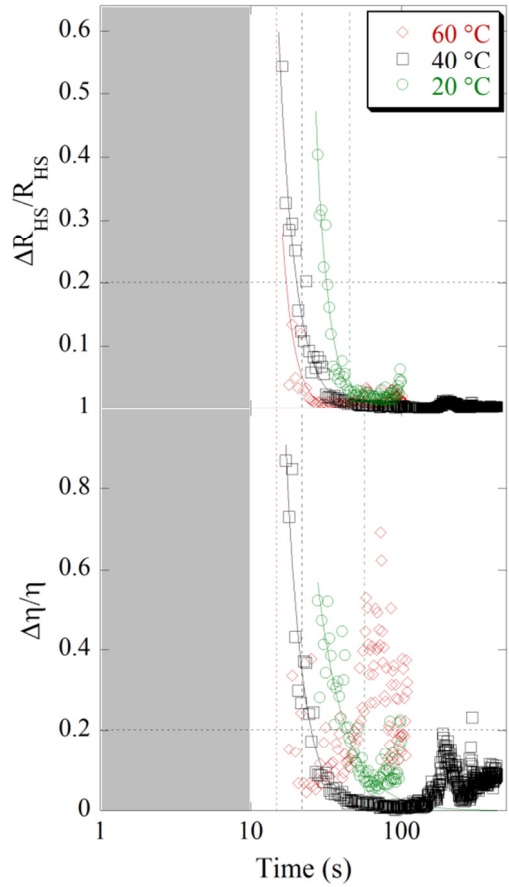

**Figure 3**: (upper panel) $\Delta R_{HS}/R_{HS}$ and (lower panel) $\Delta\eta/\eta$ as a function of the time.